\documentclass[12pt,letterpaper]{article}
\usepackage{graphicx,times,mathptm}
\usepackage{epstopdf}
\begin{document}
\baselineskip=24pt
\bibliographystyle{unsrt}

\hyphenpenalty=5000
\tolerance=1000

\vspace{3in}

\begin{center}
{\Huge Brushes of Statistically Branched Polymers}

\vspace{1in}

{ \large Daniel O. LeSher, Grant L. Metheny and Galen T. Pickett }

\end{center}
\normalsize

\newcommand {\refto} {\cite}
\newcommand {\vol} {\bf}
\long\def\omit#1{}
\newcommand {\csi} {\xi}
\newcommand {\rhot} {\rho^{\prime}}
\newcommand {\csid} {\xi^{\prime}}
\newcommand {\gsim} {\stackrel{\textstyle >}{\sim}}
\newcommand {\lsim} {\stackrel{\textstyle <}{\sim}}
\newcommand {\vo} {v_o}
\newcommand {\vot} {v_o^{\prime}}
\newcommand {\aoverd} {a^{\prime}}
\newcommand {\peq} {d^3 P_{eq}}

\begin{center}
{\it   Department of Physics and Astronomy, 
California State University Long Beach,
	1250 Bellflower Blvd., Long Beach, CA 90840 \\

}
\end{center}

\vspace{0.5in}

\begin{center}

{\large   Submitted, {\it Polymer Journal}  \today}
\end{center}

\vspace{0.5in}

\begin{center}
{\large\bf Abstract}
\end{center}

\baselineskip=24pt
We determine the distribution of free ends and the monomer insertion potential
in the strongly-stretched limit for regularly and statistically branched 
polymer brushes.
We find that the end density flattens in the limit of 
very strong
branching with a concomitant enhancement of the grafting surface
end density.
This enhancement ensures for a wide range of parameters that a 
{\it parabolic} potential profile is preserved, even for large 
positive curvature of the brush surface.
This considerably simplifies the analysis of copolymer phases.

\pagebreak
\hyphenation{mon-o-mer mon-o-mers homo-poly-mer homo-poly-mers co-poly-mer
	co-poly-mers an-i-so-trop-ic macro-molecules den-dri-mers
	Den-dri-mer}

\section{Introduction}
Geometrically branched molecules (small-molecule dendrimers
\refto{discovery} 
 and 
dendritic polymers \refto {dendrimer_polymer_review}) 
are seeing progressively sophisticated 
applications  \refto{dendrimer_review_2,dendrimer_review_1},
particularly as biomedical and drug-delivery agents
\refto{dendritic_polymer_medicine}.
The proliferation of free ends (possibly functionalized) and
a rich design space in the interior of these single molecules
makes them ideal as engineering platforms.

This work is particularly focused on determining the collective 
structure of large collections of dendritic 
or hyper-branched \refto{hyperbranched}, 
polymers end-grafted on a surface
\refto{brush}.
Even ordinary, linear polymers gain interesting properties when they 
are crowded together onto a surface by one end.
In making an analogy between the Edwards single-chain free energy and
an electrostatic system, Semenov \refto{semenov}, and subsequently
Milner, Witten, and Cates \refto{mwc} (in making an analogy to the classical
mechanics of a particle falling in an external potential), determined that
the monomer chemical potential took on a universal profile - the
``parabolic'' potential (and density profile, when the polymer brush was
solvated).
Monodispersity of the chains is a very powerful constraint producing this
behavior, as is the unique  self-consistent potential thus determined.
When the chains are grafted on a convex surface \refto{marko}
the parabolic potential profile is no longer the self-consistent solution
as such an {\it ansatz} would require monomers to overfill space near the
grafting surface (so that the self-consistently determined distribution
of free ends would be {\it negative} in some regions).
A considerably more complex analysis is required to determine the 
self-consistent potential in the presence of regions with zero end-density 
(so-called ``dead'' or ``exclusion'' zones).

However, when the polymers in the brush
 are branched (star-like \refto{katya} 
or dendritic \refto{galen_forrest}-\refto{galen_zook})
the architectural increase 
of free ends counteracts the ``overfilling'' effect, and it can 
occur that non-zero end-densities are achieved even for 
the extreme case of a single dendritic polymer in a good solvent.
Here \refto{galen_zook}, the absence of dead-zones and the consequent 
parabolic potential and therefore monomer density causes 
a single dendritic polymer to have a dense core of monomers that 
decreases monotonically to its edge in good solvent.
This is in contrast to the original prediction \refto{degennes_dendrimer}
that the tips of the dendritic polymer would all extend to the
same spherical surface producing a large, characteristic ``hollow core''
which had eluded detection in early simulations \refto{filled_core}.
The monodispersity of the hyper-branched polymer is again the culprit, as
a massive free-energy degeneracy is a consequence of the parabolic
potential  (and is the cause of the isochronous behavior of the harmonic
oscillator \refto{goldstein}).

In this work, we generalize a previous theory 
\refto{galen_forrest,dendrimer_copoly} as in Ref.~\refto{galen_zook} in a
``continuous branching'' model for the dendrimer brush.  
Vastly different architectures (regularly branched, randomly branched,
and various power-law schemes of branching) can all be handled in the
same theory with few assumptions and fitting parameters.
The parabolic shape of the insertion potential survives the
continuous branching treatment, and is the exact analytic solution for
an unreasonably large class of polymer brush architectures.

The paper is organized as follows.  First, we introduce the strong 
stretching self-consistent theory for continuously branched brushes, 
then we analyze the end-distributions for various architectures.  Finally we
discuss the implications for the near-universal stabilization of the
parabolic profile by branching, and make our conclusions.

\omit{
Dendritic molecules have received much attention since their 
invention \refto{discovery},
partly because of their intriguing architecture, but more importantly
for their promise in producing materials of decidedly interesting
properties \refto{dendrimer_review_2,dendrimer_review_1}.
Single dendrimers can be synthesized with very extremely regular
architecture, and are capable of forming thermally tunable complexes
as drug-delivery agents \refto{thermal_tune}.
Their well-characterized size and relative stiffness makes them suited
to forming complex two-dimensionally packed arrays that can then be
decorated or etched to produce nanoscopically patterned surfaces 
\refto{patterns}.
And, the geometric proliferation of chain tips, each capable of being
tagged with a bio-specific functionality, makes them an ideal material
for engineering smart surfaces \refto{darrel_tips}.
It should be pointed out that most naturally occurring biologically
relevant molecules are either lightly branched (the three-armed fatty-acids)
or linear (phospholipids, DNA, and proteins), and it is therefore
entirely possible that hyperbranching in and of itself can grant
unusual single-molecules properties with no natural analog.

Related to these dendrimers are dendritic polymers, where the 
branching points are connected by flexible spacer polymers of a controllable
molecular weight and composition.
In some respects, these dendritic polymers resemble 
polymer stars \refto{daoud}, with
a general splaying of their many free arms away from a dense core.
Indeed, miktoarm stars with an unequal number of $A$ and $B$ arms have been
predicted to drastically reshape the $AB$ diblock copolymer 
phase diagram \refto{milner_stars},
and hyperbranched dendritic polymers are predicted to have the same
effect \refto{fredrickson,dendrimer_copoly}.
The microphases are consistently skewed toward having the component
with the most branches on the exterior of a curved surface,
with the most pronounced effects occurring for stars with, for example, 
a single $A$ type arm and many $B$ arms.
Similarly, when the $A$ block is a linear flexible homopolymer, and the
$B$ block is a $G$-generation dendritic polymer, the phase diagram
is skewed considerably toward keeping the branched block on the
exterior of the cylindrical and spherical domains.
For example, it was found that for compositionally symmetric ``tadpole''
copolymers \refto{tadpole}, the lamellar 
phase is stable when the branched block 
is $G1-5$, while the cylindrical phase obtains for $G6-8$, and the spherical
phase occurs for $G9$ and more branched \refto{dendrimer_copoly}.

The purpose of the present work is to extend earlier treatments of this system
which had been limited to this most dramatic case where
the $A$ and $B$ blocks of the copolymer are both dendrimers of independent
generation.
Thus, we investigate the strong-segregation limit for $GA-GB$ block 
copolymers, and determine the phase boundaries between the ``classical''
block copolymer phases.
In the lamellar phase (L), each of the branched blocks stretches
away from the flat $AB$ interface, producing a layered material as in 
Figure~1.
There are two cylindrical phases, one (CA) with the $A$ material forming
a cylindrical core with the $B$ phase forming a continuous matrix in 
which the $A$ cores are hexagonally packed.
The second cylindrical phase (CB) has the $B$ material confined to the cores,
while the $A$ material forms the continuous matrix.
Likewise there are two spherical phases, (SA) and (SB) in which one material 
is confined to spherical domains packed on a bcc lattice, while the
other material forms a continuous matrix.
Denoting $\phi_A$ and $\phi_B$ the overall compositions of the single
dendriblock copolymers, the phase boundaries will be controlled by the
overall composition and the generations of the $A$ and $B$
blocks, $\phi_B(G_A,G_B)$.
The transitions between these phases will always follow the sequence
(SA) - (CA) - (L) - (CB) - (SB) as $\phi_B$ is increased.
The calculations we make are in the spherical/cylindrical unit cell 
approximation, and rule out  a priori exotic bicontinuous
phases.

The first set of calculations we employ involve the Alexander-de Gennes
approximation \refto{fredrickson,alexander,degennes}.
We assume that all of the $A$ tips and the $B$ tips reside 
on the same surfaces.  
In the (L) phase, this results in the $A$ tips being segregated to a single
surface extending away from the $AB$ interface, and similarly for the $B$
tips.
In the (C) phases, the outer block tips all reside on a cylindrical surface
enclosing the core cylinder, while the inner tips
are brought toward the center axis of the core,
and similarly for the (S) phases.
It should be noted that the (L) phase thus resembles back-to-back brushes
of dendritic molecules, while the (C) phase resembles the 
conformation of single dendrimer-comb copolymers, where dendrimers are
grafted to a single flexible backbone chain at regular lengths 
\refto{dendrimer_review_2}.
The (S) phase resembles the single-molecule conformation of an individual
dendrimer molecule.
Thus, the Alexander approximation can be seen as being similar to
the de Gennes and Hervet ansatz for a ``hollow-core'' 
dendrimer \refto{degennes_dendrimer}.
The ``filled-core'' picture of Muthukumar and 
Lescansec \refto{filled_core}, however, seems
to be theoretically and experimentally a more sound description.
Indeed, allowing the tips of the internal and external dendrimer to occupy
the entire lamellar/cylindrical/spherical domain turns out to be a more
realistic assumption.
In the strong-segregation limit, relaxing the confinement of the
chain free ends is achieved in the so-called ``classical limit'' 
\refto{semenov,mwc} of
the Edwards self-consistent field.

The paper is organized as follows.  We initially describe the
Alexander-de Gennes calculation, and then the classical path approximation.
Then, the phase diagrams for the dendrimer-dendrimer 
copolymer will be developed.
The results herein rest on the stong-segregation limit, the robustness
of which can be determined in a simple Random Phase Approximation
(RPA) calculation.
Finally, our conclusions will be offered.
}

\section{Model}
We use $n$ as a ``chemical index'' marking the fewest monomers required from
a given monomer to arrive at a non-grafted free end.
Thus, each free tip has a chemical index of $0$, and the unique grafted monomer
has a chemical index of $N$ (analogous to the overall
molecular weight for a linear polymer).
We denote the branching profile, $f(n)$ to stand for the number of statistically
equivalent monomers with the chemical index $n$.
For an unbranched polymer, $f(n)=1$, while for a regular 
(double tip-splitting) dendritic polymer 
of generation $G$, 
$f(n)$ is an exponentially decreasoing step function with $f(0) = 2^G$ and discontinuities at 
$n=1/G, 2/G ... (G-1)/G$.  This branching profile has been used to
analyze brushes of dendritic polymers \refto{galen_forrest} and 
copolymers of dendritic polymers \refto{dendrimer_copoly} and brushes
of star polymers \refto{katya}.
We can model polymers, however, with continuous branching profiles, such as
\begin{equation}
\label{geometric}
f(n) = 2^{-nG+G} \equiv e^{-b n}
\end{equation}
as in \refto{galen_zook} where the conformation of a single dendrimer
was considered.  
The ``branching index'', $b$ can be chosen so that we model polymers with
$G$ generations and a junction functionality of 3 modeling tip-splitting
dendrimers ... but any positive value of $b$ makes physical sense.
This ``continuously branched'' polymer model has significant analytic advantages
and loses only some of the (admittedly interesting) self-similar details of
the exact free-end density.
We will consider another class of branching profiles:
\begin{equation}
\label{powerlaw}
f(n) = \left[\frac{1 - \alpha n/N}{1-\alpha}\right]^b
\end{equation}
modeling polymers which are branched so that the average number of 
statistically equivalent monomers at any chemical index 
decreases as a power law from the free tips.
Here, $\alpha$ is a measure of how strongly branched the polymers are, with the degree of branching diverging at $\alpha \rightarrow 1$.
These two scenarios are depicted in Figure~\ref{figure1} (top panel). 

At any rate, we consider a set of polymers characterized by a maximal 
chemical index $N$, of monomers of volume $a^3$ with the $n=N$ monomer 
irreversibly grafted to a flat surface with a grafting density of $\sigma$
chains per unit area.
We consider a melt brush thus formed so that the overall height of the 
grafted layer is consistent with the incompressibility of the monomers:
\begin{equation}
\frac{h}{\sigma} = a^3 \int_0^N dn f(n).
\end{equation}
If $\sigma$ is large enough, then the chains in the brush will be considerably
stretched from the grafting surface. 
As in Figure~\ref{figure1}, the $n=N$ monomer is located at $z=0$, and 
the $f(0)$ equivalent chain tips are located at $z=z_o < h$.
The single-chain Edwards free energy of this chain is
\begin{equation}
F_{single} = \int_0^N dn f(n) \left[ \frac{1}{2 a^2} \left|\frac{dz}{dn}\right|^2 +
  a^3 P(z(n)) \right]
\label{single}
\end{equation}
Under conditions in which $F_{single} >> 1$ (in natural energy units of
$k_B T$), fluctuations of the chain conformations around the
configuration which minimizes Eq.~\ref{single} are negligible, and all
statistical averages can be calculated from the configurations minimizing 
the single chain free energy itself.
Here, the first term counts the Gaussian elastic energy of $f(n)$ equivalent
chains each consisting of $dn$ monomers and stretched a distance $dz$.
The second term counts the energy required to evacuate a volume $a^3$ for
each monomer in these test chains, at a cost in free energy of $P(z)a^3$.
Thus, $P(z)$ is the monomer pressure in the layer, created by the crowding and 
the conformations of all of the other chains.

The minimization of Eq.~\ref{single} is effected through the usual calculus
of variations, easily recognizable if the kernel of the integral is
interpreted as the Lagrangian of a particle with a time-dependent mass in 
a gravitational field $-P(z)$:
\begin{equation}
L[z;\frac{dz}{dn}] = f(n) \left[ \frac{1}{2 a^2} |\frac{dz}{dn}|^2 +
  a^3 P(z(n)) \right],
\end{equation}
with the momentum conjugate to $z(n)$ given by
\begin{equation}
p = \frac{\partial}{\partial dz/dn} L = \frac{f(n)}{a^2} \frac{dz}{dn},
\end{equation}
and the generalized force given by 
\begin{equation}
f = \frac{\partial L}{\partial z} = f(n) a^2 \frac{d}{dz} P(z(n)).
\end{equation}
Thus, the Euler-Lagrange equation governing the minimzation of $F_{single}$ is
\begin{equation}
\frac{d}{dn} p = f 
\end{equation}
or
\begin{equation}
\frac{d}{dn} f(n) \frac{dz}{dn} = f(n) a^5 \frac{dP}{dz}.
\label{eqmot}
\end{equation}
The physical initial condition on this equation of motion is that 
\begin{equation}
\left. \frac{dz}{dn}\right|_{n=0} =0,
\end{equation} 
thus requiring that there be no tension on the free chain end.

The pressure field, $P$, can be calculated self-consistently from the 
solutions to Eq.~\ref{eqmot}.
First, we need that $P$ satisfies an isochronous property.
Following the time-dependent mass analogy, a particle of mass $f(t)$ released at rest 
from a position $z_o$ must hit the grafting surface $z=0$ when $n=N$.
This condition must obtain from every position with a non-zero density of 
free ends in the layer:
\begin{equation}
\label{sc1}
N = \int_o^N dn = \int_0^{z_o} dz \frac{1}{dz/dn}.
\end{equation}
Also, given such a $P$ it must be possible to create a layer completely
filled with monomers at every $z$ without overfilling space for the melt brush we consider here.  The
common formulation of this condition is that the height density of free
tips $d \sigma / dz$ be postive:
\begin{equation}
\label{sc2}
\Phi(z) \equiv 1 = \int_z^h dz_o \frac{d \sigma (z_o)}{dz_o} \phi(z;z_o)
\end{equation}
with the volume fraction of monomers in the vicinity of $z$ for a chain with its 
$f(0)$ ends located at $z_o$:
\begin{equation}
\phi(z;z_o) = \sigma a^3 f(n) \frac{dn}{dz}.
\end{equation}
When $f(n)\equiv1$, (an ordinary polymer brush) self-consistency and the isochronous condition are satisfied 
when \refto{mwc}
\begin{equation}
P(z) = P_o (1-z^2/h^2)
\end{equation}
with $P_o = \pi^2/8  a \sigma^2$ and
\begin{equation}
\frac{d \sigma}{dz} = \sigma \frac{z/h}{\sqrt{h^2 - z^2}},
\end{equation}
remarkably compact and elegant expressions.

One appraoch to solving the self-consistency equations 
Eqs~\ref{sc1}-\ref{sc2} when $f(n) \ne 1$ is
to solve this set of integral equations for the unknown fields $P(z)$ and
$d \sigma /dz$ simultaneously and numerically.  
However, knowing that an arbitrary $f(n)$ gives the same physics as a time
dependent mass in a gravitational field, we can be confident that the 
{\em parabolic} pressure profile is a fruitful {\it ansatz}.
A time-dependent mass in a harmonic gravitational potential will have the
same falling time no matter where it is released from rest.
The equation of motion under these conditions becomes:
\begin{equation}
\label{parabolic_eq}
\frac{df}{dn} \frac{dz}{dn} + f(t) \frac{d^2 z}{dn^2} = -2 f(n) a^5 P_o z
\end{equation}
with $dz/dn|_0=0$ where the pressure scale $P_o$ still has to be 
determined self-consistently by enforcing $z(N)=0$ numerically.

To make further progress, we need to specify $f(t)$, but there  are two
general results that can be gleaned at once.
First, if the distribution of ends is non-zero for all $0<z<h$, the equation
of motion is linear in $z(n)$.
Thus, the general solution with $z(0) = z_o$ and $z'(0)=0$ satisfies:
\begin{equation}
z(n) = z_o  \xi(n),
\end{equation}
where $\xi(n)$ is a solution to Eq~\ref{parabolic_eq} with
\begin{equation}
\xi(0)=1 \mbox{   and   } \xi'(0) =0.
\end{equation}
The final condition on $\xi(n)$ is satisfied by choosing $P_o$ such that
\begin{equation}
\xi(N) = 0.
\label{po_eq}
\end{equation}
Thus, the existence and uniqueness of $\xi$ guarantees that the equal-time 
and minimum free energy conditions are satisfied.
The non-negativity of the end-distribution must be checked separately.

The second is a general degeneracy theorem.
The trajectory given by the solution to Eq~\ref{parabolic_eq} minimizes the
overall single-chain free energy, Eq~\ref{single}.  
We can very quickly show that {\em all} chain trajectories, regardless of 
the location of their free ends, $z_o$, have the {\em same single chain free
energy}.
This massive degeneracy is the source of the ``filled-core'' dendrimer 
(as all sub-branch free energies with the subbranches starting at any
position from the center out to the edge of the dendrimer have the same
energy, and therefore add equally to the statistical averages determining the
concentration profile).
The minimum single-chain free energy is
\begin{equation}
F_{min}[z(n)] = \int_0^N dn f(n) \left[ \frac{1}{2 a^2} 
\left|\frac{dz}{dn}\right|^2 +
  a^3 P_o(1- \frac{z^2}{h^2}) \right].
\end{equation}
Rewriting the first term,
\begin{equation}
\int_0^N dn f(n) \frac{1}{2 a^2} 
\left|\frac{dz}{dn}\right|^2 = 
\int_0^N dn \left[f(n) \frac{1}{2 a^2}\frac{dz}{dn}\right]\frac{dz}{dn},
\end{equation}
and integrating by parts:
\begin{equation}
=\left. z(n) f(n) \frac{1}{2 a^2}\frac{dz}{dn}  \right|_0^N -
\int_0^N dn \frac{d}{dn} \left[ f(n) \frac{1}{2 a^2}\frac{dz}{dn}\right] z =
- \int_0^N a^3 P_o \frac{z}{h^2} z,
\end{equation} using the initial condition, the isochronous condition, and
the equation of motion.
Thus, the first term exactly cancels the $z^2$ term in the minimum free 
energy yielding:
\begin{equation}
F_{min} = P_o a^3 \int_0^N f(n),
\end{equation}
so that an energy $P_o a^3$ must be paid to insert all of the $\int dn f(n)$
monomers on the branched chain.
The free energy of a chain with its free ends located at $z_o$ is thus independent
of $z_o$, so all of the chains contribute equally to the total layer energy.
Thus, the parameter $P_o$ is the overall monomer pressure to insert 
any monomer of volume $a^3$ into the brush.

It remains to calculate $P_o$ given $f(n)$ and to verify that the end-density 
is non-zero to verify the parabolic {\it ansatz} is a solution to the problem.
When the brush is composed of geometrically branched polymers, 
and Eq~\ref{geometric}
prevails, and the equation of motion becomes ($f(t) = e^{-b t}$):
\begin{equation}
-b \frac{dz}{dt} + \frac{d^2 z}{dt^2} = -2 a^5 P_o z
\end{equation}
a ``time-reversed'' linear drag harmonic oscillator.
When Eq~\ref{powerlaw} obtains (with the substitution $u = (-1+ \alpha n/N)$)
the equation of motion is
\begin{equation}
u z''(u) +  z'(u) + \frac{2 P_o a^5 N^2}{\alpha^2 h^2} u z =0,
\end{equation}
a Bessel equation.

\section{Results}
We have determined both $P_o$ and the end-density profiles by numerically
solving Eq.~\ref{po_eq} and by numerically inverting the integral equation
Eq.~\ref{sc2}.
Figure~\ref{figure2} and Figure~\ref{figure3} shows how the insertion 
potential per monomer depends on the relative degree of branching of the
polymers in the brush.
Here we refer to the ``continuous branching'' model of Ref.~\refto{galen_zook}
as a continuous version of a tip-splitting dendritic polymer and the
``power-law'' brush corresponds to choice of $f(n)$ corresponding to 
Eq.~\ref{powerlaw}, with the the power-law parameter $b=1,2,3$ 
in these calculations.  When 
$\alpha \rightarrow 1$ the polymer has more and more equivalent ends.

We have scaled $P_o$ to its value for a linear brush in the classical limit,
$ \sigma^2 a \pi^2 /8$.
In all of the cases we consider, it is apparent that the cost per monomer 
to establish the branched grafted layer decreases, and approaches zero 
in the limit of very high stretching.
This is not to say that the overall free energy per chain decreases when
the branching increases, as the energy per chain is porportional to the
number of monomers.  For chains of equivalent overall total chemical index
$N$ the cost at each $n$ is $f(n) P_o$, and thus is a rapidly increasing
function with branching.

This does imply, however, that the stretching for most of the monomers
with low $n$ (and hence high $f(n)$) is very low.  
For hyper branched chains, this implies that the vicinity of the free
ends acts essentially as densely cross-linked gel ``buoy'' tethered to 
the grafting surface by a set of very strongly stretched, but low overall
molecular weight, tethers.
The symptom of this in the free-end distributions would be a dramatic 
flattening and an enhancement of the end-density near the grafting
surface, as in the lower panel of the schematic, Fig.~\ref{figure1}.
This is exactly what we note in Fig.~\ref{figure4} and Fig.~\ref{figure5}.
The $G=4$ continuous branched layer and the $\alpha= 3/4$ linear power
law brush have essentially uniform end-density across the layer.
We note that the requirement for self-consistency is that
$\frac{d \sigma}{dn} > 0$ throughout the brush, so that the parabolic
{\it ansatz} is indeed the correct self-consistent inter-monomer interaction
potential.

\section{Discussion}
Given how successful the parabolic {\it ansatz} is, it is tempting to
jump to the erroneous conclusion that the parabolic potential is the
self-consistent potential for {\em any} choice of $f(n)$.
The two choices we have made result in nicely analytic minimum-action
trajectories, $z(n;z_o)$, but the critical feature required is that
$f(n)$ be a decreasing function of chemical index.
That is, the chains are unbranched at the grafting surface, and become more
and more branched as they continue away from the grafting surface.
If the chains started highly branched at the grafting surface, with
subchains combining into loops as one left the grafting surface so that the
chain became less and less branched toward the free ends 
then the brush will resemble a layer or Alexander-deGennes 
\refto{alexander} \refto{degennes} brushes 
of loops, with the free ends evacuated from a large region of the 
brush \refto{fredrickson}.  The self-consistent potential is then known
to be radically different from the parabolic potential here, although the
real potential still satisfies an isochronous, or monodisperse, condition.

Indeed, if we start with an unbranched set of chains, $f(n)=1$,
and alter the branching a bit by
\begin{equation}
f(n) = 1 + \epsilon(n)
\end{equation}
with $\epsilon(n)$ a small function that is very strongly peaked at $n=0$, 
it is easy to show that the surface end-density must be negative, and
space is overfilled if the parabolic potential is followed.
Essentially, we need to increase the number of chain free ends located with
$z_o$ near the full brush height, $h$, above the unbranched limit. This causes 
the number of monomers entrained into the layer to be slightly larger than they would 
have been had $\epsilon(n)=0$, with the result that building the brush from the
outer edge to the surface maintaining $\phi(z)=1$ will overfill space with 
monomers at the grafting surface.

So, if $f(n)$ is a {\em rapidly} decreasing function of $n$, each chain ``dumps''
almost all of its monomers right in the vicinity of the free end location,
$z_o$.
In this case, the molten layer resembles a set of large, dense microgels 
tethered by much lower molecular weight net of branched chains.  
The minimum-action chain trajectories become (essentially) particles at rest
for most of $n<N$, which then use their last few remaining monomers to fully
stretch back to the grafting surface.  
Thus, the layer loses its character as an anisotropic stress environment, and
becomes essentially a molten simple fluid of these free ``heads'' constrained to
exist between $0<z<h$, with a uniform density of ``heads''.
The rest of the brush is merely there to maintain the connectivity of the
chains.
The flattening of the trajectories, and concomitant flattening of the 
chain end density distributions is the result.

The enrichment of the end-density at $z=0$ has one more important consequence.
If the grafting surface is convexly curved, the tendency of the branching to enhance the surface end-density and the curvature to decrease it will cancel in favor of non negative grafting surface density ... the critical ``parabolic'' curvature will of course depend upon the properties of the grafted and branched chains, but there is the possibility that copolymer phases of branched polymers (cylindrical, and spherical in particular) will be devoid completely of free-end ``dead-zones,'' making the estimate of the free energies of the phases elementary.
Indeed, dead-zones of this type occur for energetic reasons as well, as in brushes of gradient copolymers \refto{galen_gradient}.  It is an open question as to whether branching of the gradient copolymers will close that dead-zone.

\section{Conclusion}
In the self-consistent classical path analysis with a continuous
branching profile, we have shown that the
 parabolic potential profile for hyper branched and statistically hyper
branched polymers is the self-consistent chain interaction potential.
This parabolic brush is remarkably robust to changing the chain topology
and the melt-brush thus formed is (in the limit of strong branching) a 
simple liquid of ``heads'' tethered weakly to the surface by the last
tether.  Thus, branching dramatically simplifies the structure and analysis
of these molten layers.

\pagebreak

\pagebreak

\pagebreak

{\bf \Large Figure Captions}

\begin{figure}
\includegraphics[angle=0,width=5in]{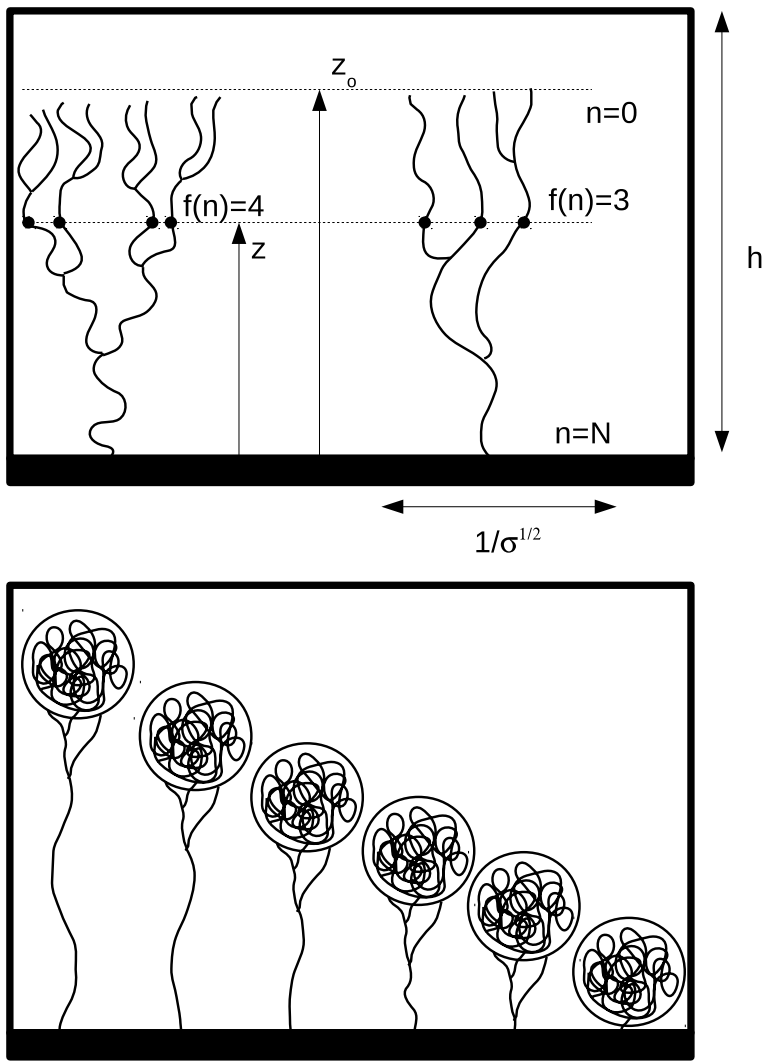}
\caption{\bf Schematic.}
\label{figure1}
\end{figure}

\noindent
\ref{figure1} {\bf Schematic.} 
The brush is constructed of polymers with chemical index running from 
$n=0$ at the free end of the polymer to $n=N$ at the grafted end.
The overall height of the layer is $h$, and $f(n)$ gives the number of 
statistically equivalent chains at a particular chemical index $n$.
Dendritic polymers have a discontinuous $n$, and we model statistically
branched polymers through continuous functions $f$.
An exponential $f$ models a traditional dendritic polymer, while 
a power law $f$ models a set of polymers in which the probablility
of a branch point is not constant as a function of chemical index $n$.

\begin{figure}
\includegraphics[angle=0,width=6.5in]{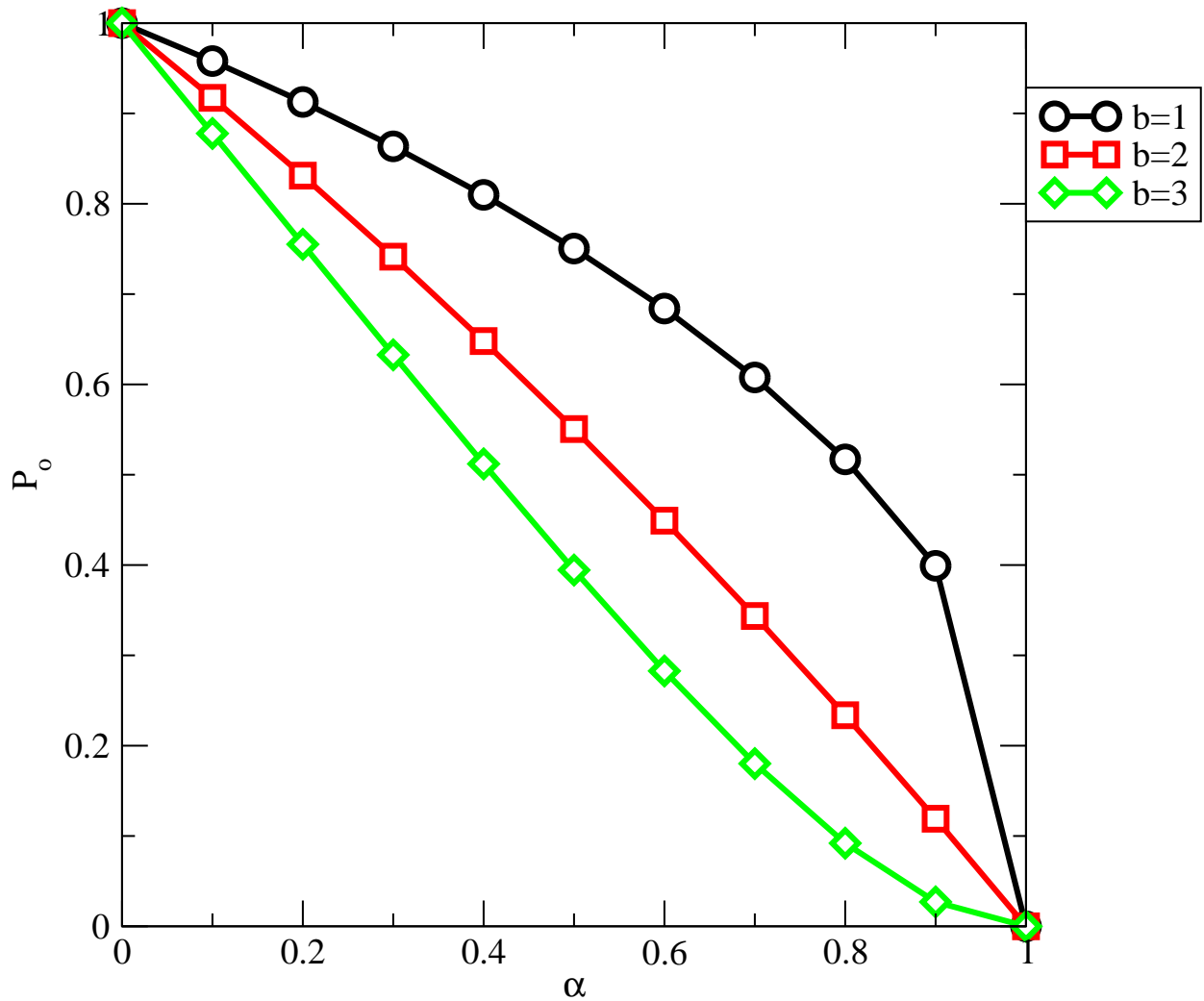}
\caption{\bf $P_o$ vs. $\alpha$ for Power Law Brush.}
\label{figure2}
\end{figure}

\noindent
\ref{figure2}  {\bf $P_o$ vs $\alpha$ for Power Law Branches}.
The pressure per monomer (scaled to the unbranched 
result, $ \sigma^2 a \pi^2/8$) as a function of the branching parameter $a$ for linear through cubic 
branching is shown.  The parameter $\alpha$ controls the 
branching magnification:
$f(n) = \left[(1- \alpha n/N)/(1 -\alpha)\right]^b$ for $b=1,2,3$.  The pressure vanishes
when $\alpha \rightarrow 1$ consistent with a highly branched polymer brush being
essentially a simple fluid of massive monomer ``heads'' connected to the 
grafting surface by single polymer strands.
While the pressure per monomer decreases, the overall free energy per chain
increases considerably as the branching increases, as there are more
and more monomers per chain.

\begin{figure}
\includegraphics[angle=0,width=6.5in]{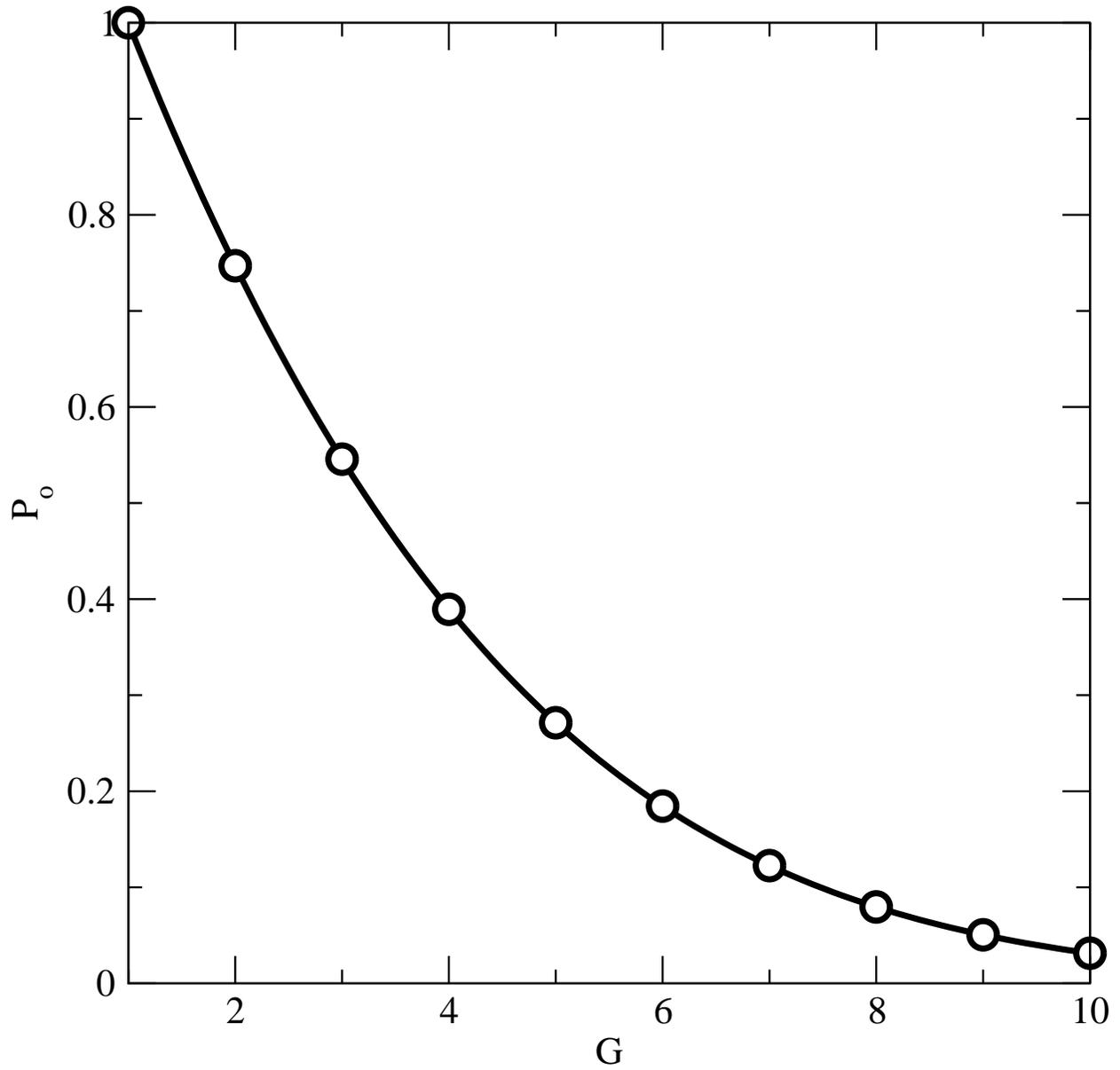}
\caption{\bf $P_o$ vs. $G$ for Continuous Branching}
\label{figure3}
\end{figure}

\noindent
\ref{figure3}  {\bf $P_o$ vs. $G$ for Continuous Branching}.
The pressure per monomer (scaled to the unbranched result, $P_o = \sigma^2 a \pi^2/8$ as a function of the effective generation of the dendritic polymer, $G$.
Here, $f(n) = 2^{G(1-n/N)}$, so that there are $2^G$ branches at the free tips
($n=0$) and a single branch at the grafting point ($n=N$).
The pressure vanishes
when $G \rightarrow \infty$ consistent with 
a highly branched polymer brush being
essentially a simple fluid of massive monomer ``heads'' connected to the 
grafting surface by single polymer strands.
While the pressure per monomer decreases, the overall free energy per chain
increases considerably as the branching increases, as there are more
and more monomers per chain.

\begin{figure}
\includegraphics[angle=0,width=6.5in]{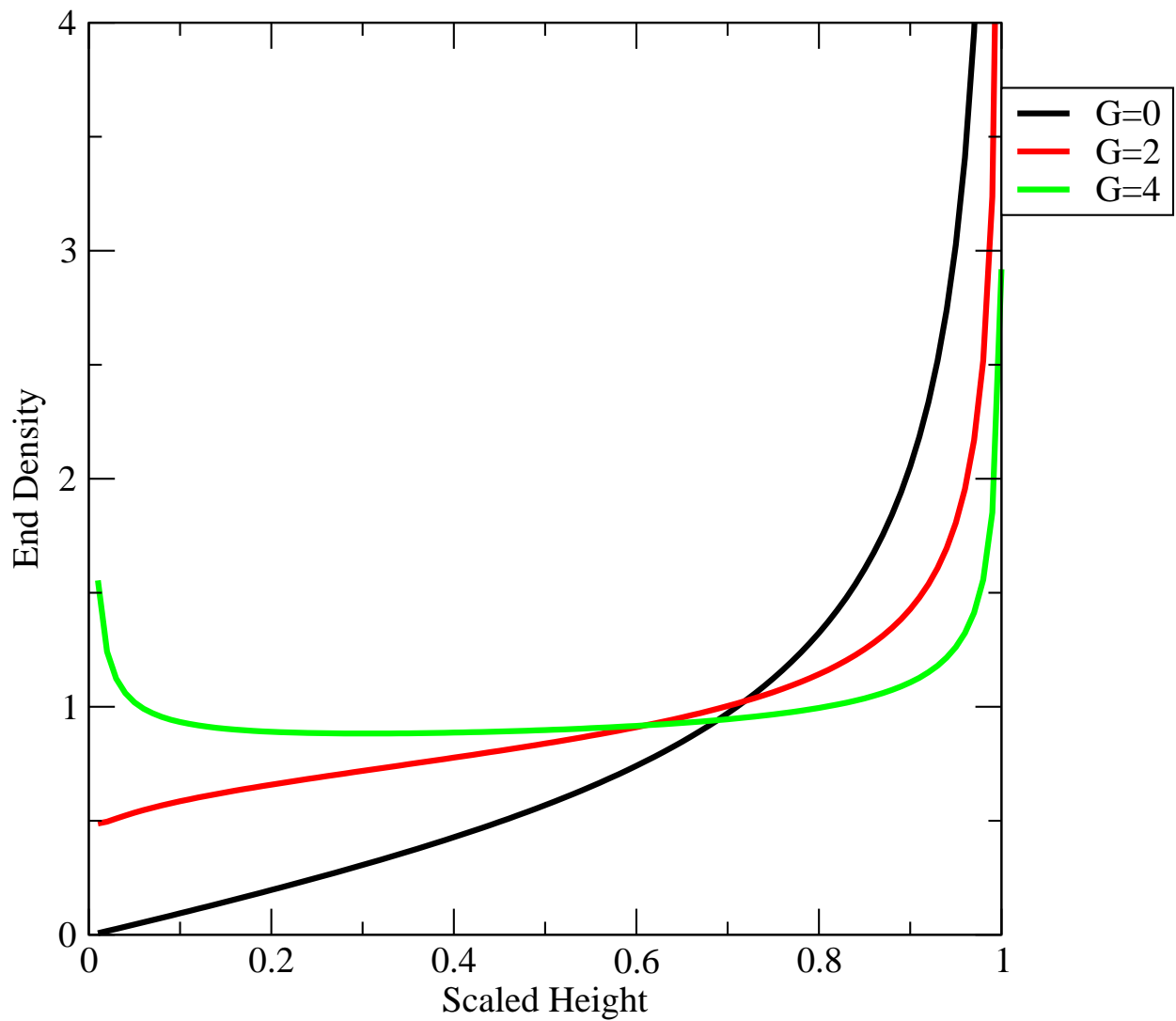}
\caption{\bf End Distributions for Continuously Branched Brushes.}
\label{figure4}
\end{figure}

\noindent
\ref{figure4}  {\bf End Distributions for Continuously Branched Brushes.}
Here we show the scaled end-density ($\frac{1}{\sigma} \frac{d \sigma(z)}{dz}$) as a function of scaled height in the brush
$z/h$.
We show various values of the effective 
dendrimer generation $G$.  
As $G$ increases, it is evident that the end-distribution becomes flatter
and flatter, and essentially uniform in the limit of high $G$.
Such a brush resembles a brush of ``buoys''.

\begin{figure}
\includegraphics[angle=0,width=6.5in]{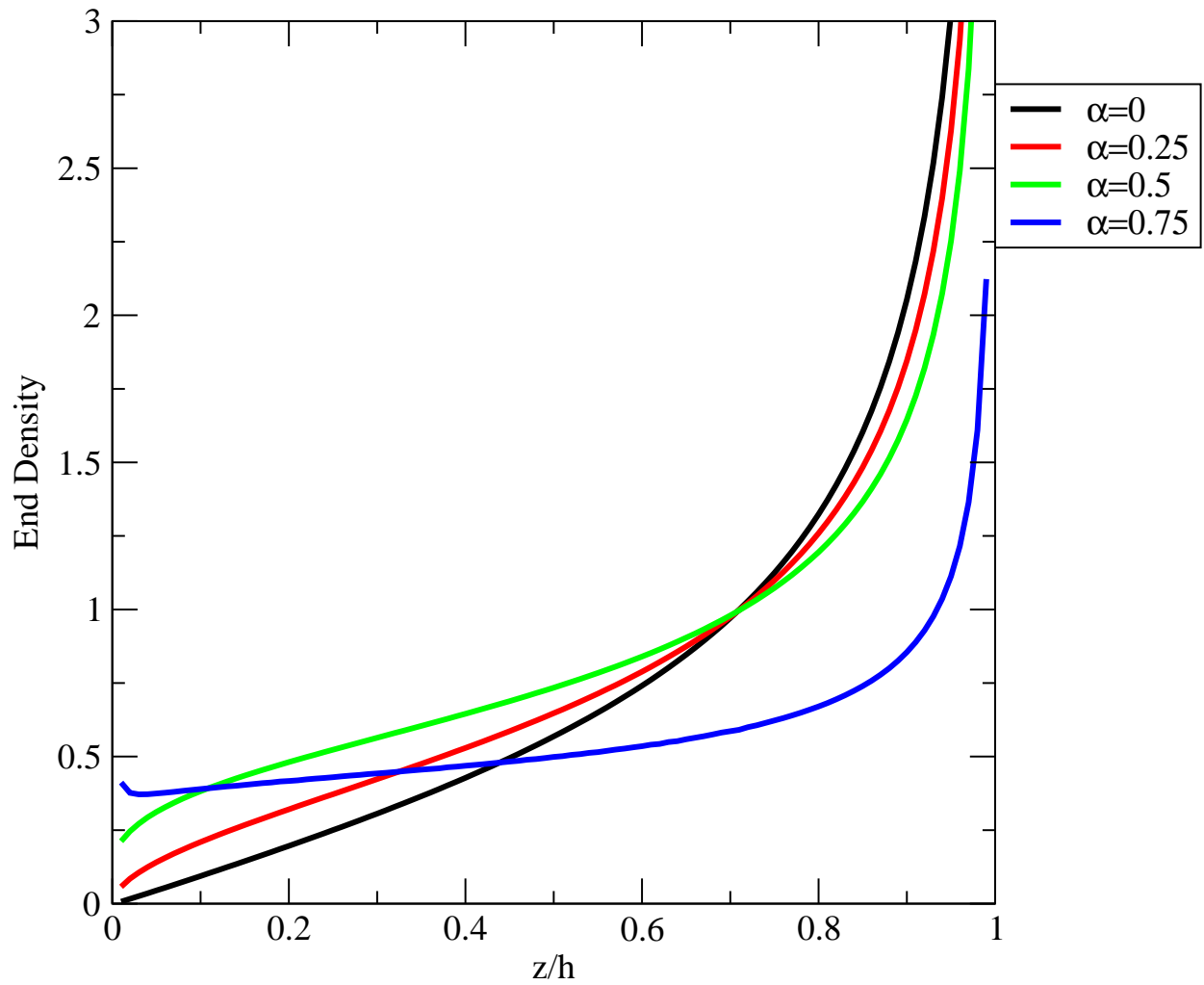}
\caption{\bf End Distributions for Power Law Branched Brushes.}
\label{figure5}
\end{figure}

\noindent
\ref{figure5}  
{\bf End Distributions for Power Law Branched Brushes}
Here we show the scaled end-density as a function of scaled height in the brush
$z/h$. 
For the linear power law ($b=1$) we show the end distribution for 
$\alpha=0, 0.25, 0.5, 0.75$.  The end-distribution is nonzero at all heights, and
becomes flatter as $\alpha$ increases.

\pagebreak

\pagebreak

\pagebreak

\pagebreak

\pagebreak

\end{document}